\documentstyle[12pt]{article}

\renewcommand{\theequation}{\thesection.\arabic{equation}}

\newcommand \beq{\begin{eqnarray}}
\newcommand \eeq{\end{eqnarray}}
\font\tennbf=cmbx12 \newfam\nbffam
\textfont\nbffam=\tennbf
\def\nbf{\fam\nbffam\tennbf}
\font\tennrm=cmr12 \newfam\nrmfam
\textfont\nrmfam=\tennrm
\def\nrm{\fam\nrmfam\tennrm}

\begin{document}


\def\bfepsilon{\mbox{\boldmath$\epsilon$}}
\def\bfgrad{\mbox{\boldmath$\grad$}}


\def\square{\hbox{{$\sqcup$}\llap{$\sqcap$}}}
\def\grad{\nabla}
\def\del{\partial}


\def\frac#1#2{{#1 \over #2}}
\def\smallfrac#1#2{{\scriptstyle {#1 \over #2}}}
\def\half{\ifinner {\scriptstyle {1 \over 2}}
   \else {1 \over 2} \fi}


\def\bra#1{\langle#1\vert}
\def\ket#1{\vert#1\rangle}


\def\simge{\mathrel{%
   \rlap{\raise 0.511ex \hbox{$>$}}{\lower 0.511ex \hbox{$\sim$}}}}
\def\simle{\mathrel{
   \rlap{\raise 0.511ex \hbox{$<$}}{\lower 0.511ex \hbox{$\sim$}}}}


\def\parenbar#1{{\null\!
   \mathop#1\limits^{\hbox{\fiverm (--)}}
   \!\null}}
\def\nunubar{\parenbar{\nu}}
\def\ppbar{\parenbar{p}}


\def\buildchar#1#2#3{{\null\!
   \mathop#1\limits^{#2}_{#3}
   \!\null}}
\def\overcirc#1{\buildchar{#1}{\circ}{}}


\def\slashchar#1{\setbox0=\hbox{$#1$}
   \dimen0=\wd0
   \setbox1=\hbox{/} \dimen1=\wd1
   \ifdim\dimen0>\dimen1
      \rlap{\hbox to \dimen0{\hfil/\hfil}}
      #1
   \else
      \rlap{\hbox to \dimen1{\hfil$#1$\hfil}}
      /
   \fi}


\def\subrightarrow#1{
  \setbox0=\hbox{
    $\displaystyle\mathop{}
    \limits_{#1}$}
  \dimen0=\wd0
  \advance \dimen0 by .5em
  \mathrel{
    \mathop{\hbox to \dimen0{\rightarrowfill}}
       \limits_{#1}}}                           


\def\real{\mathop{\rm Re}\nolimits}     
\def\imag{\mathop{\rm Im}\nolimits}     

\def\tr{\mathop{\rm tr}\nolimits}       
\def\Tr{\mathop{\rm Tr}\nolimits}       
\def\Det{\mathop{\rm Det}\nolimits}     

\def\mod{\mathop{\rm mod}\nolimits}     
\def\wrt{\mathop{\rm wrt}\nolimits}     


\def\TeV{{\rm TeV}}                     
\def\GeV{{\rm GeV}}                     
\def\MeV{{\rm MeV}}                     
\def\KeV{{\rm KeV}}                     
\def\eV{{\rm eV}}                       

\def\mb{{\rm mb}}                       
\def\mub{\hbox{$\mu$b}}                 
\def\nb{{\rm nb}}                       
\def\pb{{\rm pb}}                       



\def\journal#1#2#3#4{\ {#1}{\bf #2} ({#3})\  {#4}}


\def\AdvPhys{\journal{Adv.\ Phys.}}
\def\AnnPhys{\journal{Ann.\ Phys.}}
\def\EurophysLett{\journal{Europhys.\ Lett.}}
\def\JApplPhys{\journal{J.\ Appl.\ Phys.}}
\def\JMathPhys{\journal{J.\ Math.\ Phys.}}
\def\LettNuovoCimento{\journal{Lett.\ Nuovo Cimento}}
\def\Nature{\journal{Nature}}
\def\NPA{\journal{Nucl.\ Phys.\ {\bf A}}}
\def\NPB{\journal{Nucl.\ Phys.\ {\bf B}}}
\def\NuovoCimento{\journal{Nuovo Cimento}}
\def\Physica{\journal{Physica}}
\def\PLA{\journal{Phys.\ Lett.\ {\bf A}}}
\def\PLB{\journal{Phys.\ Lett.\ {\bf B}}}
\def\PhysRev{\journal{Phys.\ Rev.}}
\def\PRC{\journal{Phys.\ Rev.\ {\bf C}}}
\def\PRD{\journal{Phys.\ Rev.\ {\bf D}}}
\def\PRL{\journal{Phys.\ Rev.\ Lett.}}
\def\PhysRept{\journal{Phys.\ Repts.}}
\def\ProcNatlAcadSci{\journal{Proc.\ Natl.\ Acad.\ Sci.}}
\def\ProcRoySoc{\journal{Proc.\ Roy.\ Soc.\ London Ser.\ A}}
\def\RevModPhys{\journal{Rev.\ Mod.\ Phys.}}
\def\Science{\journal{Science}}
\def\SovPhysJETP{\journal{Sov.\ Phys.\ JETP}}
\def\SovPhysJETPLett{\journal{Sov.\ Phys.\ JETP Lett.}}
\def\SovJNuclPhys{\journal{Sov.\ J.\ Nucl.\ Phys.}}
\def\SovPhysDoklady{\journal{Sov.\ Phys.\ Doklady}}
\def\ZPhys{\journal{Z.\ Phys.}}
\def\ZPhysA{\journal{Z.\ Phys.\ A}}
\def\ZPhysB{\journal{Z.\ Phys.\ B}}
\def\ZPhysC{\journal{Z.\ Phys.\ C}}


\begin{titlepage}
\begin{flushright} {Saclay-T95/055}\\ {IPNO/TH 95-28}
\end{flushright}
\vspace*{0.2cm}
\begin{center}
\baselineskip=13pt {\Large NON PERTURBATIVE ASPECTS \\}
 {\Large OF SCREENING PHENOMENA\\ }
\vspace*{.4cm}
{\Large IN ABELIAN AND NON ABELIAN GAUGE THEORIES}
 \vskip0.5cm Jean-Paul
BLAIZOT\footnote{CNRS} \\ {\it Service
de Physique Th\'eorique\footnote{Laboratoire de la Direction des Sciences
de
la Mati\`ere du Commissariat \`a l'Energie Atomique}, CE-Saclay \\ 91191
Gif-sur-Yvette, France}

 and \\
 Edmond IANCU \\ {\it Division de Physique
Th\'eorique\footnote{Unit\'e de Recherche des Universit\'es Paris XI et Paris
VI associ\'ee au CNRS}, I.P.N.-Orsay\\
         91406 Orsay, France}
\end{center}

\vskip 1cm
\begin{abstract}
When computed to one-loop order in resummed perturbation theory,
the non-abelian Debye mass appears to be logarithmically
sensitive to the magnetic scale $g^2T$.  More generally,  we show
that in higher orders power-like infrared divergences
 forbid the use of perturbation theory to calculate the
corrections to Debye screening.
 A similar infrared problem occurs in the determination of the
mass-shell for the scalar propagator in 2+1-dimensional
 scalar electrodynamics. In this context,
we provide a  non-perturbative approach which solves the
infrared problems and allows for an accurate calculation of
the scalar propagator in the vicinity of the mass-shell.
\end{abstract}

\vskip 1.6cm


\end{titlepage}

\setcounter{equation}{0}
\section{Introduction}

Significant progress has been achieved during the last few
 years toward the
understanding of the infrared structure of high temperature
QCD\cite{BP90,FT90,us}. The prominent role of the soft energy scales
 $gT$ and $g^2T$ has been recognized,
and the  collective nature of the dominant behaviour at the
scale $gT$ has been properly understood.
(Here, $g\equiv g(T)$ is the coupling constant at the temperature $T$, and
we assume that $g\ll 1$ in the high-temperature, deconfined phase of QCD.)
This led to a systematic description, in classical
terms, of a variety of collective phenomena like screening,
Landau damping, or color oscillations\cite{us}. In the case of screening,
it is known that, to leading order in $g$, the electrostatic
  interactions are screened, with a screening mass
 $m_D\sim gT$, while the magnetostatic interactions are not
screened\cite{Klimov81,Weldon82}.
These properties are shared by  abelian
and non-abelian plasmas (see Refs.
\cite{Fradkin65,Silin60,QED,BIR95} for the abelian case).

Important
differences occur between abelian and non-abelian gauge theories
when  corrections to the leading order Debye screening are
considered. In the abelian case,  perturbation theory can be used
to calculate the corrections to the leading order Debye mass
(see \cite{BIR95} and references therein). In QCD,  infrared
divergences occur in such a calculation,
whose origin is the coupling of the chromoelectric field
to the unscreened magnetostatic fields
(see, e.g., Ref. \cite{RebhanBanff} for a survey
of the computations prior to 1993, and also Refs. [11--15] for more
recent calculations). For example, at one-loop order,
there is a logarithmic singularity, widely discussed
in the literature\cite{Rebhan93,BN94,Rebhan94}.
But we shall see that the difficulty is actually more serious,
since power-like infrared divergences occur
in the higher orders.

The existence of infrared divergences invalidating the perturbative
expansion of thermal QCD is well known for
the magnetostatic sector, where power-like
 infrared divergences are indeed expected in higher
order calculations of thermodynamical quantities\cite{Linde80,GPY81}.
It has long been recognized that,
because only static modes are involved,
these divergences are essentially those of an
effective {\it three-dimensional} theory\cite{Appel81}.
The divergences that we shall encounter here,
which  are also those of an effective three-dimensional theory,
 are of a slightly different nature. They occur in the
perturbative evaluation
of the polarisation tensor of the  electrostatic gluon
{\it on  the tree-level mass-shell}. Similar
 divergences are encountered in the calculation of
the quasiparticle damping rates (see, e.g., \cite{Rebhan95}
and references therein). All such divergences could be
removed by introducing an infrared cut-off $\lambda$
 in the magnetostatic sector.
 However, this is not a satisfactory solution
for at least two reasons. In QCD, there is  a common belief that
such a cut-off is indeed
generated dynamically in the form of a magnetic mass
$\lambda\sim g^2T$ \cite{GPY81,BMuller93}. But for such a value of $\lambda$
 there are infinitely many terms in
the perturbative expansion which contribute to the same order,
a situation analogous to the Linde problem\cite{Linde80}.
The second reason is that we shall identify
 similar mass-shell divergences
in the evaluation of the static scalar propagator in thermal scalar
electrodynamics (SQED). And we know that there is no magnetic mass
in abelian gauge theories\cite{Fradkin65,BIR95}.

Thus, although we expect the picture of Debye screening
to hold in higher order calculations, for reasons which will
be detailed in the next section, it appears that the corresponding
value of the screening mass cannot be computed in
 perturbation theory beyond the leading order. We are thus led
to look for a non perturbative description which allows
for the treatment of the large degeneracy of states involving
massless magnetostatic fields.  We shall propose
such a treatment for SQED, and obtain the mass-shell behaviour
of the scalar propagator without any  infrared regulator.

Our analysis relies on a non-perturbative
 approximation to the Dyson-Schwinger equations.
The method that we use, known as the  {\it gauge technique},
has been developed
originally\cite{Salam63,Delbourgo} in relation
to abelian gauge theories in four dimensions, and has
been found to be particularly convenient for the study of the
infrared structure of the propagator.
It has the advantage to preserve the correct Ward identities, and the
expected  analytical properties of the propagator.
Within this  formalism, we shall be able to determine
 the infrared behaviour of the  three dimensional
 scalar propagator. We shall find that the
 mass-shell singularity, which is  a simple pole in
leading order, turns into a  branch point,
whose location can be shown to be gauge-fixing independent.

The plan for the rest of the paper is as follows.
In section 2, we introduce the screening function, and discuss
its analytic properties. In section 3, we critically analyze the previous
computations of the Debye mass in the resummed one-loop approximation,
and show that  the loop expansion generates power-like
infrared divergences.
We identify a similar problem in the charged  sector
of SQED. In section 4, we present a non-perturbative approach
which allows us to study
  the mass-shell behaviour of the scalar propagator
in 2+1-dimensional SQED.  The last section summarizes
the conclusions.

\setcounter{equation}{0}
\section{The screening function}

In electrodynamics, the potential between two static charges
 in a medium  can be
calculated from the  electrostatic propagator in the medium, to
be referred as the {\it screening function} in what follows,
\beq\label{S}
S( x)\equiv \int_{-\infty}^{\infty} {\nrm d}x_0\,
D_{00}(x_0, {\bf x})\,=\,\int\frac{{\nrm d}^3 k}{(2\pi)^3}\,\,
\frac{ {\nrm e}^{i{\bf k}\cdot{\bf x} } }{k^2 -
\Pi_{00}(0,k)}\,,\eeq where $x=|{\nbf x}|$ is the distance between
the two static charges, $k=|{\nbf k}|=\sqrt{{\nbf k}^2}$,
 $\Pi_{00}(0,k)$ is the static ($k_0=0$) electric polarisation tensor,
and $D_{00}(0,k)$ the electrostatic propagator: $D_{00}(0,k)= 1/(k^2 -
\Pi_{00}(0,k))\equiv S(k)$. We shall need later to analytically continue
$S(k)$
to complex values of $k$. In most cases to be discussed
 in this paper, $S(k)$
will be obtained explicitly as an  even function of  $k$. When this
is not so, we shall regard
$S$ as a function of $\sqrt{k^2}$, i.e.
$S(k)=S(\sqrt{k^2})$, before doing the analytic
continuation.

A similar screening function describes the
interaction of two static color charges in a quark-gluon plasma. In
this case, however, the polarization tensor is
  gauge-dependent\cite{KK82,Nadkarni86,Rebhan93}. A fully gauge invariant
treatment of chromoelectric screening should start from a gauge invariant
object, such as the Polyakov loop. But, in
perturbation theory, the leading long
range behaviour of the correlator of two Polyakov loops is
determined by the screening function (2.1) (see Refs. \cite{BN94,Rebhan94}).
(We are not implying here that perturbation theory
correctly describes the long range behaviour
of the Polyakov loop correlator, which is presumably dominated
by glue ball intermediate states\cite{Braaten95}.)
For this reason, we shall concentrate
on this simpler object here. In fact, the long-distance behaviour
of $S(x)$ turns out to be gauge-independent. This may be expected
from general arguments \cite{KKR90}, and will be
verified explicitly.

Let us then return to eq.~(\ref{S}) which we
rewrite as
\beq\label{S1}
S( x) = \frac{1}{2\pi x}\, \int_{-\infty}^{\infty} \frac
{{\nrm d}k}{2\pi i}\,
k\,S(k)\,{\nrm e}^{ikx}\,.\eeq
In leading order, $S_0(k)=1/(k^2+m^2)$ where
\beq\label{mel}
m^2 \,=\, -
\delta \Pi_{00}(0,0) = \frac{g^2 N T^2}{3}\,\eeq
is the leading-order screening mass squared and
 $ \delta\Pi_{\mu\nu}$ is the polarization tensor in the ``hard thermal
loop'' approximation [1-5](we consider here a pure gluonic plasma). The
integral (\ref{S1}) may be computed by continuing the integrand to
complex values of $k$, and by closing the integration path in the
upper half of the complex $k$-plane.
One then picks up the contribution of the pole $k= im$, and
gets the familiar screened Coulomb potential,
$S_0(x)= {\nrm e}^{-mx}/4\pi x$.

In abelian plasmas, the higher order corrections do not change
significantly the picture. The  singularity of $S(k)$ which
controls the asymptotic behaviour of $S(x)$ remains a pole on the
imaginary axis; recently, this has been verified explicitly
up to two orders beyond the hard thermal loop approximation
\cite{BIR95}.  In such a case, we can define a screening mass
$m_D$ as the solution of \cite{Rebhan93}
\beq\label{debyemass}
m_D^2=-\Pi_{00}(0,k)\left.\right|_{k^2=-m_D^2}.
\eeq
This self-consistent equation admits also a meaningful perturbative
solution\cite{BIR95}.

Higher order corrections to $\Pi_{00}(0,k)$ play a more dramatic
role in QCD, where they alter the nature of the singularities of
$S(k)$. In order to discuss these corrections, it is convenient to
remark that, in leading order, they
 can be considered
 as loop corrections in the effective
 three-dimensional theory  obtained after integrating the
non-static loops with static external lines
(see \cite{Nadkarni83,Braaten94} and references therein).  At
the order of interest, and in Coulomb or covariant gauges,
the corresponding Euclidean action reads
\beq\label{Seff}
S_{eff} = \int {\nrm d}^3 x \,{\nrm Tr}\,\Bigl ( \frac{1}{2}\,F_{ij}^2
+ (D_i A_0)^2 + m^2 A_0^2+\frac{1}{2\zeta}(\del_i A_i)^2+
\del_i\bar\eta D_i\eta) \Bigr )\,
,\eeq where
$D_i=\del_i - ig\sqrt{T} A_i$, $F_{ij}=[D_i, D_j]/(ig\sqrt{T})$,
 $m$ is the leading-order electric mass  from eq.~(\ref{mel}),
$\zeta$ is the gauge fixing parameter, and $\eta$ and $\bar\eta$
are the ghost fields. The effective theory describes interacting
static and long-wavelength ($k \simle gT$)  fields. The
magnetostatic gauge
 fields $A_i^a({\nbf x})$
and the electrostatic field $A_0^a({\nbf x})$ are, up
to normalizations, the zero-frequency components of the original gluonic
fields.

In the effective theory, $A_0^a$ enters as a
massive scalar field, whose propagator is the screening function.
We shall write
\beq\label{Sk1}
S^{-1}(k)\equiv k^2+m^2+\Sigma(k)
\eeq
with $\Sigma(k)$ denoting the self-energy corrections in the
 three-dimensional theory,
i.e. we set  $\Pi_{00}(0,k) = -m^2 -\Sigma(k)$.
It is easy to see that all the Feynman diagrams contributing to $\Sigma(k)$
are analytic in $k^2$ for small momenta. Indeed,
in any such diagram, one can choose the independent loop momenta
 so that the external momentum ${\nbf k}$
 flows only along the massive propagators. These can be expanded
 with respect to ${\nbf k}$,  when
$|k|\ll m$. In the resulting expression, the external
momentum appears then only in the numerator, and  rotational symmetry
ensures that only the terms with even powers of ${\nbf{k}}$
survive the angular integration.

One may regard the effective action (\ref{Seff}) as the Euclidean
version of a Minkovskian action in 2+1 dimensions.
 From this point of view, one expects $S(k)$ to be
analytic in the whole complex $k$-plane, except
 on the imaginary axis. We shall assume that this property
indeed holds and write
\beq\label{Leh}
S(k)\,=\,\int_0^{\infty}{\nrm d}\omega\,\frac{\rho(\omega)}
{\omega^2 + k^2}\,.\eeq
The spectral density $\rho(\omega)$ may be calculated from the
discontinuity of $S(k)$ accross the imaginary axis:
\beq\label{rho}
\rho(\omega)\,\equiv\,\frac{2\omega}{\pi}\,{\nrm
Im}\,S(k=i(\omega+i\epsilon))\,\eeq
with $\omega$ real. Because $S$ is
an even function of $k$,
$\rho(\omega)$ is an even function of $\omega$ (this is easily verified
by noticing that $S(k)$ is real for $k$ real, and applying the Schwartz
reflexion principle: $S^*(k)=S(k^*)$). Thus, only the positive values of
$\omega$ are needed to represent $S(k)$. At  leading order,
$S(k) = 1/(k^2+m^2)$ and
$\rho(\omega)=2m\delta
(\omega^2-m^2)$. We define
\beq\label{tildesig}
\tilde\Sigma(\omega)\equiv\Sigma(k=i(\omega+i\epsilon))
\eeq
where $\omega$ is real. Then
\beq\label{rhosig}
\rho(\omega)= -\,\frac{2\omega}{\pi}\,\frac {{\nrm Im}\,
\tilde\Sigma (\omega)} {\Bigl [ \omega^2 - m^2 - {\nrm Re}\tilde
\Sigma (\omega)\Bigr ]^2 \,+\,
\left[{\nrm Im} \,\tilde\Sigma (\omega)\right ]^2}\,.\eeq
In a Minkovskian theory, one expects $\rho(\omega)$ to be positive
in a physical gauge. However, what is meant by a physical
gauge is not the same in the present (2+1)-dimensional
problem  and in the original (3+1)-dimensional one. For the original
problem at finite temperature, one can choose, as a physical gauge,
 the strict Coulomb gauge, that is, the limit $\zeta\to 0$
in eq.~(\ref{Seff}).
In the  Minkovskian theory in 2+1 dimensions, this does not  correspond
 to a Coulomb gauge, but rather to a Landau gauge which
 involves unphysical degrees of freedom.
Since these latter do not give positive definite contributions to
the spectral density,  $\rho(\omega)$ is not then necessarily
 positive for the theory defined by the effective action eq.~(\ref{Seff}).

Since the three-dimensional theory (\ref{Seff}) is
superrenormalisable,   $\Sigma(k)/k^2\to 0$ as
$k\to
\infty$.  It follows then from eq.(2.6) that, as $|k|\to\infty$,
$S(k)\simeq 1/k^2$. This property allows us to derive a sum rule for
$\rho$. First we note that, owing to the asymptotic property of $S$
just mentioned, we have
\beq
\oint\frac{{\nrm d}k}{2\pi i} \,k\, S(k)=1\,,
\eeq
where the contour is a circle at infinity in the complex $k$-plane.
Then we replace in this equation $S(k)$ by its expression
(\ref{Leh}) in terms of the spectral function. The contributions of
the two poles at $k=\pm i\omega$ give
\beq\label{sum-rule}
\int_0^\infty {\nrm d}\omega\,\rho(\omega)=1.
\eeq
 We shall
see later that, in some approximations, $S(k)$ may have poles away
from the imaginary axis, which are not accounted for by the
spectral function $\rho(\omega)$. We
shall  argue later that such poles are unphysical, but it is
nevertheless useful to keep track of them by writing
\beq
S(k)=\int_0^\infty{\rm d}\omega\frac{\rho(\omega)}{\omega^2+k^2}+
\sum_i\left( \frac{a_i}{k^2-k_i^2} + \frac{a_i^*}{k^2-k_i^{*2}}\right).
\eeq
The sum rule
 is then modified into
\beq\label{sum-rule2}
\int_0^\infty {\nrm d}\,\omega\rho(\omega)+\sum_i (a_i +a_i^*)=1.
\eeq

We conclude this section by summarizing the analytic properties
that we expect the screening function
$S(k)$ to satisfy.  We have given
arguments suggesting that $S(k)$ is analytic in the complex
$k$-plane, with singularities on the imaginary axis. Furthermore,
in most gauges,
$S(k)$  is
 analytic in $k^2$ for small $|k|$, i.e. $k\ll gT$. We shall verify  that
these properties are satisfied by the approximate $S(k)$ that we
shall obtain. We shall find that $S(k)$  has branch  cuts along the
imaginary axis, starting at $k=\pm im^*$ (see Fig.~1). The branch
point dominates the asymptotic behaviour of $S(x)$. According
to eqs.~(\ref{S1}) and (\ref{Leh}), we can write
\beq\label{Scut}
S(x)=\frac{1}{4\pi x}\,\int_{m^*}^{\infty} {\nrm d}\omega\,
{\nrm e}^{-\omega x}\,\rho(\omega)\,,\eeq
which expresses the screening function as the Laplace transform of the
spectral density. At large $x$, $S(x)\sim f(x)e^{-m^*x}$,
and we shall verify that $m^*$ is gauge-fixing
independent.

The previous analyticity arguments, which are sufficient
to establish the exponential fall off of the screening function
at large distances, may
become invalid in some particular gauges. In particular, in
the  temporal axial gauge, the asymptotic fall-off of
$S(x)$ was found to be a power law
 rather than an exponential\cite{BaierK94,Wong94}.
It is likely however that this peculiar behaviour is a gauge
artifact (see also Ref. \cite{Rebhan94}). Indeed,
the temporal axial gauge is known to lead to specific
difficulties  in the imaginary-time
formalism\cite{GPY81}. It prevents in particular the power
counting arguments leading to the effective three dimensional
action (\ref{Seff}).

\setcounter{equation}{0}
\section{Corrections to Debye screening are non-perturbative}

In perturbation theory, we expect the dominant singularity of
$S(k)$ to remain close to the leading-order pole at $k=im$.
Thus, the determination of the Debye mass involves the calculation
  of $\Sigma(k)$ for $k\sim im$. Since
 $\Sigma(k=im)$ is infrared singular in perturbation theory,
this leads to difficulties whose physical origin is analyzed
in this section. We shall be led finally to the conclusion
that perturbation theory cannot be used to estimate the corrections
to the leading-order Debye mass.

\subsection{The polarisation tensor at next to leading
order}

The one-loop graph contributing to $\Sigma$ is
displayed in Fig. 2. It is readily evaluated as
\beq\label{Sigma1l}
\Sigma(k;m)=
- g^2 NT
 \int {{\nrm d}^D q \over (2\pi)^D} \,\frac{(2k_i +q_i)(2k_j+q_j)}
{({\nbf q+k})^2+m^2}\,D_{ij}^0(0,{\nbf q})\,,\eeq
where ($\hat q_i= q_i/q$)
\beq\label{Dij}
 D_{ij}^0(0,{\nbf q})=\frac{\delta_{ij}-\hat
q_i\hat q_j}{{\nbf q}^2}+ \zeta \,\frac{\hat q_i \hat q_j
}{{\nbf q}^2} \eeq
is the free propagator for magnetic gluons.
We are  using here, and throughout,  dimensional continuation in order
to regularise the {\it ultraviolet} (mass) divergences.
 After computing the integrals,
 no UV singularity will actually subsist in the limit $D\to 3$.
It is important to keep in mind that the limit $D\to 3$ will
always be taken {\it before} discussing the infrared structure of
the integrals. After a simple rearrangement, eq.~(\ref{Sigma1l})
takes the form
\beq\label{sig1l}
\Sigma(k;m)= g^2 NT
 \int {{\nrm d}^D q \over (2\pi)^D} \Biggl\{
\frac{1}{{\nbf q}^2+m^2} 
+\frac{2}{{\nbf q}^2}\frac{m^2-{\nbf k}^2}{({\nbf q+k})^2+m^2}\\ \nonumber
+\left({\zeta} -1\right)
\left({\nbf k}^2+m^2\right)\, \frac{{\nbf q}\cdot \left({\nbf q+2k}\right)}
{{{\nbf q}^4}\left(({\nbf q+k})^2+m^2\right)} \Biggr\}\,,
\eeq
and gives, after an  elementary integration,
\beq\label{SIG*}
\Sigma(k;m)= \alpha m
\left\{ \frac{2(m^2-{k}^2)}{m k}
 {\nrm arctan}\frac{k}{m} \,+\, \left({\zeta} -2\right)\right\},\eeq
 where $\alpha \equiv {g^2N T}/{4\pi}$ has the dimension of a
mass.

The function (\ref{SIG*}) has logarithmic branch points at
$k=\pm im$. The origin of this singularity may be seen  on
eq.~(\ref{sig1l}): as  $k^2 \to -m^2$, the dominant contribution
to the integral comes from the small $q$ region, and when $k=\pm im$ the
integral in fact diverges. To understand physically what happens,
it is convenient to do a Wick rotation. We have, for
real $\omega$ (recall eq.~(\ref{tildesig})),
\beq\label{tilsig}
\tilde\Sigma(\omega;m)\,=\, \alpha m
\left\{ \frac{(m^2 + \omega^2)}{ m \omega}
 {\ln}\,\frac{m+ \omega +i\epsilon}
{m -\omega -i\epsilon} \,+\,\left({\zeta} -2\right)\right\},\eeq
from which one gets
\beq\label{imsig}
{\nrm Im}\,\tilde\Sigma(\omega;m)\,=\,\pi \alpha\,\frac
{\omega^2 + m^2}{\omega}\,\theta(\omega^2-m^2)\,.\eeq
This imaginary part
 is proportional to $\Phi(\omega)$, the invariant phase-space
for the decay of a particle of energy $\omega$ into a
particle of mass $m<\omega$ and a massless particle. This
is easily computed  in the rest frame of the
decaying particle as
\beq\label{PhS}
\Phi(\omega)= \int\frac{{\rm d}^2q}{(2\pi)^2}
\frac{1}{2\epsilon_{q} \,2q}\, \delta(\omega-q-\epsilon_{q})
=\frac{1}{8\pi\,\omega}\theta(\omega^2-m^2)\eeq
where $\epsilon_q=\sqrt{q^2+m^2}$. Note that
this 2-dimensional phase space does not vanish when $\omega\to m_+$,
in contrast to the 3-dimensional one. This behaviour of the phase-space
factor is responsible for the infrared divergences to be discussed
further in section 3.3.

Another noteworthy feature of eq.~(\ref{imsig}) is that
 ${\nrm Im}\,\tilde\Sigma(\omega;m)>0$ for $\omega >0$, whereas with
our conventions we would
expect the opposite sign. The sign of ${\nrm
Im}\,\tilde\Sigma$ is related to that of the spectral
function according to eq.~(\ref{rhosig}), so that,
 in the one-loop approximation, $\rho(\omega)$
is negative for all momenta $\omega>m$.
 In particular, from  eq.~(\ref{rhosig}) and the asymptotic
 form of the one-loop self-energy (\ref{tilsig}),
namely $\tilde\Sigma(\omega\to \infty)\simeq i\pi \alpha\omega$,
one obtains, for large $\omega$,
\beq\label{asymp0}
\rho(\omega)\sim -2\frac{\alpha}{\omega^2}.
\eeq
Thus, as alluded to after eq.~(\ref{rhosig}), in the present covariant gauge
 $\rho(\omega)$ cannot be regarded as a physical
spectral density.

Because the poles at $k=\pm im$ of the unperturbed propagator
coincide with the branch points in the self-energy, which
furthermore diverges in these points, the equation
(\ref{debyemass}) cannot be
used to calculate perturbatively
the correction to the Debye mass. In fact the
analytic structure of the propagator
$S(k)=1/(k^2 + m^2 +\Sigma(k))$ is very different from that of the
unperturbed one. It has  branch points at $k=\pm im$
and, besides, a set of four simple poles at $k = \pm a \pm ib$,
where the real numbers
$a$ and $b$ are  gauge-dependent \cite{Rebhan94}.
 To leading order in $\alpha$, the values of
$a$ and $b$ are given by $a=\alpha (\pi-\theta)$,
$b-m\approx(\alpha/2)[\zeta-2+\ln(4m^2/(a^2+(b-m)^2))]$, with
$\theta=\arctan(a/(b-m))$.
It is instructive to follow the trajectory of these poles in the
complex $k$-plane, as a function of the dimensionless parameter
$\alpha/m\sim g$. In order to do so, we rewrite the inverse
propagator as
\beq
S^{-1}(k)=m^2\left\{ 1+x^2+\frac{\alpha}{m}\,f(x)\right\}
\eeq
where $x\equiv k/m$ and $f(x)\equiv \Sigma/(\alpha m)$. For small
coupling, the poles behave as indicated above. When the coupling
increases, they follow the trajectories displayed in Fig.~3
(for the gauge $\zeta=2$). There
exists a critical coupling at which the poles become real. Beyond
that, one of the pole flows toward $m$, the other being equal to
$\alpha \pi$. We note that the latter regime corresponds to the
small mass regime, which is attained here at strong coupling. The
pole at $k=\alpha \pi$ is the tachyonic pole already identified in
the studies of the massless theory\cite{Jackiw81,Appel81}.

This analytic structure of the one-loop propagator, which
contradicts the expected properties of $S(k)$, leads
to unphysical properties.
 Indeed, the relative magnitudes of $b$ and $m$, which determine
the asymptotic behaviour of $S(x)$, depends on the gauge.
If $b>m$, the long-range behaviour of the screening function remains
dominated  by the logarithmic singularity at $z=im$ so that
 for $x\to \infty$,
$S(x)\approx  f(x)\,{\nrm e}^{-mx}/x$.
However, since the spectral function is negative, the
pre-exponential factor
$f(x)$  is strictly {\it negative} (see eq.~(\ref{Scut})),
in contrast to the  leading order result $f_0=1/4\pi$.
If now $b<m$, the pole contributions dominate, and  the screening
function oscillates asymptotically.
These changes of regimes for small changes in the parameters
are physically  not satisfactory.

As we have mentioned after eq.~(\ref{SIG*}), the singularities of
the integral (\ref{sig1l}) are determined by the small $q$ region. They are
therefore very sensitive to the small momentum behaviour of the magnetic
gluon propagator. By allowing for a small gluon mass $\lambda$,
i.e. replacing
 $ 1/{\nbf q}^2 \,\to \, 1/({\nbf q}^2 +\lambda^2)$
 in the integral (\ref{sig1l}),
one separates the mass-shell of the scalar particle and the
threshold for gluon emission, and this removes the infrared
divergences. A simple calculation gives then\cite{Rebhan93} (for
$\lambda \ll m$)
\beq\label{siglam}
\Sigma_{\lambda}(k;m)= \alpha m \left\{ \frac{2(m^2- k^2)}{ m k }
 {\arctan}\,\frac{k}{m+\lambda}\, -1\, + \left({\zeta}
-1\right)
\,\frac{m^2+k^2}{(m+\lambda)^2 +k^2}\right\}.\eeq
In eq.~(\ref{siglam}), the branch point has now
moved to $m+\lambda$. In perturbation theory, $S(k)$
has a pole at $k=i(m+\delta m)$ where $\delta m
\equiv \Sigma_{\lambda}(k=im;m)/2m\,\approx \alpha\ln
(2m/\lambda)$.
This perturbative analysis
is consistent as long as the new pole does not
move back into the cut, that is, as long as $\alpha\ln
(2m/\lambda)<\lambda$. However, if $\lambda\sim g^2T$,
which is the order of magnitude expected for the magnetic mass,
$\alpha\ln (2m/\lambda)\sim g^2T\ln(1/g)\gg\lambda$,
 and one gets an inconsistency.

One way to keep the pole separated from the branch cut is to change
$m$ in eq. (\ref{siglam}) into $m_D$. That puts the branch point at
$m_D+\lambda$, whatever the value of $m_D$ is.  The Debye mass $m_D$ is
then obtained by solving  self-consistently the equation
\cite{Rebhan94}
\beq\label{Rebhan}
m_D^2 \,=\,m^2 \,+\,\Sigma_{\lambda}(k=im_D\,;m_D)\,.\eeq
The   pole at $k= im_D$ remains below the branch point
at $k = i(m_D+\lambda)$, so that it  controls the long range behaviour of the
screening function.
 However the sign problem alluded to earlier is not solved. The sign  of
$S(x)$  at large distances is determined by the residue at the pole.
 A simple calculation gives
\beq\label{res}
 S(x)\sim_{x\to \infty} {2\lambda \over 2\lambda + (\zeta
-3)\alpha}\,\frac {{\nrm e}^{-m_D x}}{4\pi x}\,,\eeq
the first fraction being the residue just mentioned.
Thus,  in Feynman's gauge for instance, $S(x)$ becomes negative if $\lambda
<\alpha$.  Note that, in perturbation theory, one would
expect the residue to be close to unity; the above formula
shows that this only happens if the infrared cut-off is large enough,
 i.e. $\lambda \gg \alpha$,
which is not to be expected. The above procedure leading to
eq.~(\ref{Rebhan}) is an attempt to go beyond perturbation theory,
which ignores, however, all the vertex corrections. The fact that the
latter may be important is suggested by the
 non-perturbative character of the residue.
Nevertheless, we shall see in section 4.4 that, in the presence
of a magnetic mass, the result (\ref{Rebhan}) remains correct
 even when vertex corrections are taken into account.

In closing this subsection, let us mention that, for
nonvanishing $\lambda$, the same correction to the Debye mass
as obtained above, i.e.  $\delta m \approx \alpha\ln
(2m/\lambda)$, can be deduced from the long range behaviour
of the correlator of two Polyakov
loops\cite{BN94,Rebhan94}.
Note however that the two calculations are not independent
since, in perturbation theory, the asymptotic behaviours
of the Polyakov loop correlator and of the screening function
involve the same integrals.

\subsection{ A similar problem in scalar QED}

In the high temperature limit, and at leading order in $e$, the static and
long-wavelength  ($k\simle eT$) correlation functions of scalar QED can be
calculated from the effective three-dimensional euclidean action
\cite{BIR95}
\beq\label{Leff}
S_{eff} = \int {\nrm d}^3 x \,\biggl ( \frac{1}{4}\,F_{ij}^2
+\frac{1}{2\zeta}(\del_i A_i)^2
+\frac{1}{2} \, (\del_i A_0)^2+\frac{1}{2} \,m_{el}^2\,A_0^2
 \nonumber\\
+(D_i\phi)^\dagger(D_i\phi) + m^2 \phi^\dagger\phi +
+e^2T A_0^2 \phi^\dagger\phi
+\frac{\lambda}{4}(\phi^\dagger\phi)^2\biggr )\,.\eeq
The new notations here are as follows: $\phi({\nbf x})$ is the complex
scalar field, 
 $m_{el}^2 = e^2 T^2/3$ is the leading order electric mass,
and $m^2 = e^2 T^2/4$ is the charged particle thermal mass.
We are only interested here in the  mass-shell singularities
associated with the interaction between the  charged particles and
the massless transverse photons. We shall therefore  restrict
ourselves to the sector of (\ref{Leff}) which describes these
 interactions:
\beq\label{Sred}
S_{eff} = \int {\nrm d}^3 x \,\biggl ( \frac{1}{4}\,F_{ij}^2
+\frac{1}{2\zeta}(\del_i A_i)^2
+(D_i\phi)^\dagger(D_i\phi) + m^2 \phi^\dagger\phi \biggr)\,.\eeq
There is an obvious similarity between this action and the
corresponding one for the hot QCD plasma, eq.~(\ref{Seff}): they
both describe massive charged particles ($A_0^a$ in QCD and
 $\phi$ in SQED) in interaction with massless gauge
fields. In particular, to one loop order,
 the scalar self-energy $\Sigma(k;m)$ is also given by eq.~(\ref{SIG*})
\cite{BIR95}, and most of the discussion  in section 3.1
applies to SQED as well. Of course,
essential differences persist between  the two theories in
 the dynamics of the  gauge fields themselves. In
particular, in SQED,  the transverse photons remain massless to
all orders in perturbation theory\cite{Fradkin65,BIR95}, so that
one cannot invoke anymore a magnetic mass to regularise the
mass-shell singularities, as we did in eq.~(\ref{siglam}) for QCD.

 The massless version of the theory (\ref{Sred}) has been
studied extensively\cite{Appel81}. In this case the one-loop scalar
self-energy generates a tachyonic pole at $k\sim \alpha$. This
can be seen
 from the expression (\ref{SIG*}): in the limit
$m/k \to 0$, $\Sigma(k)=  -\pi \alpha k$, and the inverse
propagator $S(k) = k^2 -\pi \alpha k$ vanishes at $k=\pi \alpha$.
The infrared behaviour is improved by the
resummation of
 the one-loop  polarization tensor in the internal {\it photon} line
 in Fig. 2. To see that, consider the photon polarization tensor
 $\Pi_{ij}(0,q)\equiv  (\delta_{ij} - \hat q_i \hat
q_j)\Pi_T(0,q)$.
 At one-loop order in the effective theory we have\cite{Kraemmer94,BIR95}
 \beq\label{iiop}
\Pi_{T}^{(1)}(0, {q})\,=\,\alpha \,m\,
\left\{\frac{4 m^2+ q^2}{2qm}\,{\nrm
arctan}\,\frac{q}{2m}\,-1\right\}.\eeq In the massless limit $m/q
\to 0$, $\Pi_{T}^{(1)}(0, {q})\,=\,\pi \alpha\, q/4$ is
{\it linear} in $q$, so that it dominates over the contribution
of  the bare inverse propagator at small momentum, that is, as $q\to
0$,
$D_{ij}(0,q)\propto 1/q$.

Although this softening of the photon propagator does not solve entirely
the infrared problem, it is clear that the loop insertions in the
internal photon lines do play an important  role in the massless
case. This is not so in the massive theory.  The reason is
that, when $m\ne 0$, and to all orders in perturbation theory,
the polarization operator is expected
to vanish at least as $q^2$ when $q\to 0$\cite{BIR95}
 (in particular, $\Pi_{T}^{(1)}(0, {q})\,\sim
\,(\alpha /6 m)\,q^2$ as $q \to 0$); thus, in the massive theory,
the low momentum
behaviour of  the resummed photon propagator is not different from
that of the bare propagator.

 \subsection{The need for a non perturbative treatment}

The infrared divergences that arise in the one-loop calculation
signal, in fact, a
breakdown of perturbation theory. There exists indeed
 an infinite number of multi-loop diagrams contributing to
$\Sigma (k)$ which become infrared singular as the external momentum
approaches the tree-level mass-shell, i.e. when $k^2\to -m^2$. The
one-loop diagram represented in Fig.~2
is logarithmically divergent as $k\to im$. Consider the
  two loop diagram
of Fig.~4a. Its complete infrared behaviour is calculated explicitly
in the Appendix, but the leading terms can be obtained  by simple
power counting.   The most divergent contribution
is in the integral
\beq\label{PC}
\int\frac{ {\nrm d}^3{q}\,
{\nrm d}^3{p} }{ (q^2+\lambda^2)(p^2+\lambda^2)({\nbf k\cdot
q})^2({\nbf k\cdot(p+q)}) }
\eeq
where we have added a small mass to the
 photon in order to facilitate the power
counting.  The integral over $q$ is linearly divergent as
$\lambda\to 0$. The same result holds for the diagram in Fig.~4b,
which involves vertex corrections, and it can be verified
that the leading divergences of the diagrams 4a and 4b do not
mutually cancel. A
similar power counting argument can be  extended to all
Feynman diagrams involving no correction to the magnetic photon
line,  such as the one displayed in Fig. 5. The result of power
counting is that, close to the mass shell, a
$n$-loop graph ($n\ge 2$)   diverges like $
(\alpha/ \lambda)^{n-1} \,$, up to powers of $\ln (\alpha/\lambda)$.

Physically, the origin of the infrared
divergences is the degeneracy
 between the mass-shell of the charged particle and the threshold
for the emission of $n$  ($n\ge 1$) massless transverse
photons. Then, the determination
of the mass shell requires  solving the theory
in the subspace of these degenerate states. Naively,
one would expect the coupling
to two or more photons --- which brings in more powers of
$g$ --- to be less important than the coupling
to  a single photon. This is what happens in 3+1 dimensional
electrodynamics. In the present case,
the low dimensionality of the space-time amplifies the effects
of the degeneracy, in such a way that the couplings to any
number of photons become equally important.

Similar divergences arise in QCD as well.
Some of the relevant diagrams are actually the same
 as in SQED (e.g., Figs.~4a,b and Fig.~5),
the scalar line in these diagrams being interpreted
as an  electrostatic gluon. Besides, there exist
 new divergent graphs involving the
self-interactions of the magnetic gluons
(see Fig. 4c for an example). The same
 power counting as above leads again to the conclusion
 that $n$-loop diagram diverge as $(\alpha/ \lambda)^{n-1}$.
One may argue that the infrared divergences
are cured by the dynamical generation of a magnetic mass
$\lambda$. However, for $\lambda \sim g^2 T \sim \alpha$, as commonly
expected\cite{Linde80,GPY81}, all the aforementioned
 diagrams contribute to the same order in $g$.

To summarize, the analysis of this section suggests that
non perturbative methods are necessary in order to determine
the correct mass shell behaviour. Such a method will be
presented in the next section for the case of SQED.

\setcounter{equation}{0}
\section{An integral equation for the spectral density}

We present now an approximate, but non-perturbative,
solution of the Dyson-Schwinger equation of scalar
electrodynamics, which provides
the behaviour of the scalar propagator near the mass shell.
To this aim, we establish a linear
 integral equation for the spectral density
$\rho (\omega)$ using the so-called {\it gauge
technique}\cite{Salam63}. This equation performs
a partial resummation of the most infrared singular diagrams in
a gauge-invariant  way, i.e. by respecting the  Ward identities.
When applied to  four-dimensional abelian gauge theories,
it  provides the correct
 mass-shell behaviour for charged particles\cite{Delbourgo}.

\subsection{The quenched approximation}

The four skeleton diagrams which enter the  Dyson-Schwinger
 equation for the  scalar propagator  are displayed in Fig. 6.
It can be verified by power counting
that, at a given order in $e$, the most singular diagrams
are obtained  from the perturbative expansion
of the first graph, Fig.~6.a, where we keep the
photon propagator at the tree-level
(see Fig.~7). These are precisely the diagrams discussed
in section 3.3.
Thus, the  Dyson-Schwinger equation that we wish to solve,
and to which we refer as the quenched approximation, is
\beq\label{DS0}
\Sigma(k)\,=\,
- e^2T \int {{\nrm d}^D q \over (2\pi)^D} \,
(2k_i + q_i)\,
D^{0}_{ij}({\nbf q})\,\Gamma_j({\nbf k +q, k})\,S({\nbf k+q})\,.\eeq
In this equation,
 $S$ and $\Gamma_i$ are the full propagator and vertex,
related by the Ward identity:
\beq\label{Ward}
q_i\, \Gamma_i({\nbf k, k+q})\,=\,
 S^{-1}({\nbf k+q})-  S^{-1}({\nbf k})\,.\eeq
The most general  vertex function which is consistent
with this identity and which is free of kinematical singularities
is of the form\cite{BallChiu}
\beq\label{gammafull}
 \Gamma_i({\nbf k, k+q})\,=\,\frac{2k_i+q_i}{({\nbf 2k+q})\cdot {\nbf q}}
\,\Bigl(S^{-1}({\nbf k+q})-  S^{-1}({\nbf k})\Bigr)\,+\,
A({\delta_{ij}-\hat
q_i\hat q_j})k_j\,,\eeq
where $A\equiv A(k^2, q^2, {\nbf k}\cdot{\nbf q})$ is an unknown
scalar function. According to the usual terminology in
the literature, we shall refer to the two terms in the r.h.s.
as the longitudinal and transverse pieces of the vertex function,
respectively. The second term, proportional to $A$, is indeed
transverse  to the photon momentum ${\nbf q}$. However,
 the first term is {\it not} parallel
 to  ${\nbf q}$: it involves a non-trivial
transverse piece which is completely determined by the
Ward identity (see also eq.~(\ref{gammat}) below).
At leading order, $S^{-1}({\nbf k})={\nbf k}^2+m^2$,
$A=0$, and  $\Gamma_i ({\nbf k, k+q})$ given by
eq.~(\ref{gammafull}) reduces to the tree-level vertex, $2k_i + q_i$.

As $q\to 0$, eq.~(\ref{gammafull}) must reproduce the differential
form of the Ward identity, namely
\beq\label{warddiff}
 \Gamma_i({\nbf k, k})\,=\,\frac{\del S^{-1}}{\del k_i}\,=\,
2k_i\frac{\del S^{-1}}{\del k^2}\,.\eeq
Together with  eq.~(\ref{gammafull}), this implies that $A\to 0$
as $q\to 0$, for any $k$. This suggests that
only the longitudinal vertex becomes singular
 as $q\to 0$ and $k^2\to -m_D^2$;  the corresponding
 singularities are, of course, those
of the 2-point function. This conclusion is supported by
perturbative calculations\cite{BallChiu}.
  Thus, in order to get the leading mass-shell
behaviour, we need only consider the longitudinal vertex.
That is, we make the following ansatz for $\Gamma_i\,$:
\beq\label{gamma1}
 \Gamma_i({\nbf k, k+q})\,=\,\frac{2k_i+q_i}{({\nbf 2k+q})\cdot {\nbf q}}
\,\Bigl(S^{-1}({\nbf k+q})-  S^{-1}({\nbf k})\Bigr)\,.\eeq
As anticipated, this has a component transverse to ${\nbf q}$:
\beq\label{gammat}
({\delta_{ij}-\hat q_i\hat q_j})
 \Gamma_j({\nbf k, k+q})\,=\,
\frac{2({\delta_{ij}-\hat q_i\hat q_j})k_j}
{({\nbf 2k+q})\cdot {\nbf q}}
\,\Bigl(S^{-1}({\nbf k+q})-  S^{-1}({\nbf k})\Bigr)\,.\eeq
It is this component which
couples to the physical, transverse piece of the photon
propagator (\ref{Dij}). The longitudinal
piece of (\ref{gamma1})  couples only to the gauge degrees
of freedom of $D^{0}_{ij}({\nbf q})$.

With the ansatz (\ref{gamma1}) for  $\Gamma_i\,$,
  the Dyson-Schwinger equation (\ref{DS0}) becomes
\beq\label{DS01}
\Sigma(k)\,=\,
- e^2T \int {{\nrm d}^D q \over (2\pi)^D} \,
D^{0}_{ij}({\nbf q})\,
\frac{(2k_i + q_i)(2k_j+q_j)}{({\nbf 2k+q})\cdot {\nbf q}}\,
\,\Bigl(1\,-\,  S^{-1}({\nbf k})S({\nbf k+q})\Bigr)\,.\eeq
 We are eventually interested in the
self-energy close to the mass-shell, $\Sigma(k\to im_D)$.
By definition, $S^{-1}( k=im_D)=0$, so that
the  term proportional to $S^{-1}(k)$
 in the r.h.s. of eq.~(\ref{DS01}) does not contribute
to $\Sigma(k=im_D)$, unless the integral  over $q$
diverges. Since this integral is indeed potentially
divergent, we introduce a  regulator, in the form
of a small photon mass. In doing so, it is important for what follows
to keep explicit the distinction between the physical
 and the unphysical states in the photon propagator.
The general structure of the exact magnetostatic propagator in
finite-temperature four-dimensional SQED is\cite{Fradkin65,BIR95}:
\beq\label{photon}
 D_{ij}(0,{\nbf q})=\frac{\delta_{ij}-\hat
q_i\hat q_j}{{\nbf q}^2+\Pi_T(0,q)}+ \zeta \,\frac{\hat q_i \hat q_j
}{{\nbf q}^2}, \eeq
with  $\Pi_T(0,q)\equiv \Pi_{ii}(0,q)/2$. We have already mentioned,
in section 3.2, that $\Pi_T(0,q)\propto q^2$ as $q\to 0$.
However, as a convenient IR regularisation, we give the transverse
photons a mass and set temporarly $\Pi_T(0,q)=\lambda_M$ ($\lambda_M\ll m$).
We also regularise the gauge sector of $D_{ij}$
by replacing  $ 1/{\nbf q}^2 \,\to \, 1/({\nbf q}^2 +\lambda^2)$
in the second term of eq.~(\ref{photon}). Both $\zeta$ and $\lambda$
should disappear in the evaluation of physical quantities.
Then  we get from
eq.~(\ref{DS01}) $\Sigma =\Sigma_L + \zeta\delta \Sigma$ where
\beq\label{DSL}
\Sigma_L(k)\,=\,
- e^2T \int {{\nrm d}^D q \over (2\pi)^D} \,\frac{1}{q^2+\lambda_M^2}\,
\frac{4({\nbf k}\cdot \hat{\nbf q})^2}
{({\nbf 2k+q})\cdot {\nbf q}}\,
\,\Bigl(1\,-\,  S^{-1}({\nbf k})S({\nbf k+q})\Bigr)\,,\eeq
is the self-energy in the Landau gauge,  while
\beq\label{zetasig1}
\delta\Sigma(k)= - e^2T
\int \frac{{\nrm d }^D{ q}} {(2\pi)^D}
\,\frac{{\nbf q\cdot(2k+q)}}{(q^2+\lambda^2)^2} \,
 \left(1-S^{-1}({\nbf k})S({\nbf k+q})\right)\,, \eeq
is the gauge dependent part of $\Sigma$.
Note that the above equation for $\delta\Sigma$ is actually independent
of the ansatz used for the vertex $\Gamma_i$, since it follows
directly from eq.~(\ref{DS0}) and the Ward identity
(\ref{Ward}).

Consider now eq.~(\ref{DSL}) in the on-shell limit $k\to im_D$.
 As long as we keep the infrared  regulator
$\lambda_M \ne 0$, the integral is convergent
and the  term proportional to $S^{-1}(k)$ vanishes
on the mass-shell. Thus
\beq\label{on-shell}
\Sigma(k=im_D)\,=\,- e^2T \int {{\nrm d}^D q \over
(2\pi)^D} \,\frac{1}{q^2+\lambda_M^2}\,
\frac{4({\nbf k}\cdot \hat{\nbf q})^2}
{({\nbf 2k+q})\cdot {\nbf q}}\bigg|
_{k=im_D}\,\simeq \,2\alpha m_D \ln \frac{2m_D}{\lambda_M}\,.\eeq
On the other hand, in the physical limit
  $\lambda_M \to 0$, not only does the estimate
(\ref{on-shell}) become logarithmically divergent,
but  the integral multiplying
$S^{-1}( k)$ also diverges on mass-shell, since,
in this limit,  it is proportional to
\beq \int\frac{ {\nrm d}^3{q}}
{q^2(q^2+2{\nbf k\cdot q})
(q^2+2{\nbf k\cdot q}+\Sigma({\nbf k+q})-\Sigma(k))}\,.\eeq
Thus the use of an infrared regulator does not allow us
to explore in a simple way the behaviour of the scalar propagator
near the mass shell. For this, more powerful technics are needed,
such as that developed in the next subsection.

Consider finally the gauge-dependent part of the self-energy,
i.e., $\delta\Sigma$, eq.~(\ref{zetasig1}).  Since
\beq \int \frac{{\nrm d }^D{ q}} {(2\pi)^D}
\,\frac{{\nbf q\cdot(2k+q)}}{(q^2+\lambda^2)^2} =
 \int \frac{{\nrm d }^D{ q}} {(2\pi)^D}
\,\frac{q^2}{(q^2+\lambda^2)^2} = - \frac{3\lambda}{8\pi}\,,\eeq
which vanishes as $\lambda \to 0$, the r.h.s. of eq.~(\ref{zetasig1})
reduces to
\beq\label{zetasig}
\delta\Sigma(k)=  e^2T\, S^{-1}({\nbf k})
\int \frac{{\nrm d }^D{ q}} {(2\pi)^D}
\,\frac{{\nbf q\cdot(2k+q)}}{(q^2+\lambda^2)^2} \,
S({\nbf k+q}).\eeq
The usefulness of $\lambda$ appears when considering
the on-shell limit of this equation. If we set $\lambda =0$,
then the $q$-integral diverges on the mass-shell, and the
limit $\delta\Sigma(k\to im_D)$  is not obvious.
But if we keep $\lambda\ne 0$, we obtain
$\delta\Sigma(k=im_D)=0$.
This guarantees  the gauge-independence
of the mass-shell, since  the equation  $S^{-1}( k=im_D)=0$
reduces then to  $m^{2}_D=m^2+\Sigma_L(k^2=-m^2_D)$.
This procedure for taking the on-shell limit in the presence
of an infrared regulator has been proposed
in Ref. \cite{Rebhan93}, at the level of the resummed one-loop
approximation. (See also Refs. \cite{Schiff92,Rebhan93b,BP92}
for a similar problem in the computation of the damping
rates.)


\subsection{The integral equation}

We shall transform now the  Dyson-Schwinger equation (\ref{DS01})
into an integral equation for the spectral density $\rho(\omega)$.
In this way, we take automatically into account the
expected analytical properties of both the propagator and the vertex.
First we put eq.~(\ref{DS0}) in the form
\beq\label{DS1}
1\,=\, (k^2 + m^2)S({\nbf k}) -
e^2T \int {{\nrm d}^D q \over (2\pi)^D} \,(2k_i + q_i)
D^{0}_{ij}({\nbf q})\,S({\nbf k+q})\,
\Gamma_j({\nbf k +q, k})\,S({\nbf k}).\eeq
Then we use  the spectral representation (\ref{Leh}) of the propagator
to rewrite the ansatz (\ref{gamma1}) for the vertex function as
\beq\label{Gamma}
 S({\nbf k})\, \Gamma_i({\nbf k, k+q})\, S({\nbf k+q}) =
\int_0^{\infty}{\nrm d}\omega\,
\frac{(2k_i + q_i) {\rho(\omega)}} {(\omega^2 +{\nbf k}^2)
\left (\omega^2 +({\nbf k+q})^2 \right)}\,.\eeq
 When eqs.~(\ref{Leh}) and  (\ref{Gamma}) are used in (\ref{DS1}),
the following equation for $\rho$ is obtained\cite{Salam63,Delbourgo}:
\beq\label{DS2}
1\,=\,\int_0^{\infty}{\nrm d}s\,\frac{\rho(s)}
{s^2 + {k}^2 }\,\Bigl[k^2 \,+\, m^2\,+\,
\Sigma(k;s)\Bigr]\,,\eeq
where $\Sigma(k;s)$ is the one-loop
expression (\ref{SIG*})
 in which the mass $m$ is replaced by $s$. That is,
\beq\label{sigomega}
\Sigma(k;s)= \alpha s \left\{ \frac{(s^2- k^2)}{i s k
 } {\ln}\,\frac{s+ik}{s -ik}\, +\, \zeta-2\right\},\eeq
where $\alpha = e^2T/4\pi$.

As it stands, eq.(\ref{DS2}) is not easy to solve.
To make progress, we do a Wick
rotation $k\to i(\omega+i\epsilon)\,$ to time like
momenta.  The integral  equation becomes then
\beq\label{DS3}
1\,=\,\int_0^{\infty}{\nrm d}s\,\frac{\rho(s)}
{s^2 - { \omega}^2 - i\epsilon}\,\Bigl[-\omega^2 \,+\, m^2\,+\,
\tilde \Sigma(\omega;s)\Bigr]\,.\eeq
By taking the imaginary part
of (\ref{DS2}) we obtain an equation which is linear in $\rho$ and
 homogeneous: \beq\label{r1}
\frac{\pi}{2\omega}\,\rho(\omega)\,\Bigl[\omega^2 - m^2 -
{\nrm Re}\, \tilde \Sigma(\omega;\omega)\Bigr]
\,=\,\int_0^{\infty}{\nrm d}s\,\frac{\rho(s)}
{s^2 - {\omega}^2}\,{\nrm Im}\,\tilde \Sigma(\omega;s)\,.\eeq

At this point, all what we have done applies to 3+1 dimensional SQED as
well. It is interesting to see what happens in this case to enlighten the
difference with the 2+1-dimensional case. In 3+1 dimensions,
\beq\label{3+1}
{\nrm Im}\,\tilde\Sigma(\omega;s)=\frac{(3-\zeta)e^2}{16\pi}
\,\frac{\omega^2+s^2}{\omega^2}\,(\omega^2-s^2)\,\theta(\omega^2-s^2).
\eeq
The real part, ${\nrm Re}\,\tilde\Sigma (\omega;\omega)$,
 is regular after UV renormalisation
and it can be combined with the
parameter $m^2$ in the l.h.s. of eq. (\ref{r1}) to  define
the physical mass (which we continue to denote by the symbol $m$ for
simplicity).
The equation for the spectral density is then\cite{Delbourgo}
\beq
(\omega^2-m^2)\rho(\omega)=\frac{(\zeta-3)e^2\,\omega}
{8\pi^2}\int_m^\omega{\nrm
d}s\,\rho(s)\left(
1+\frac{s^2}{\omega^2}\right).
\eeq
For $\omega\sim m$, this equation can be transformed into a
differential  equation
which is easily solved. To do that, set $F(\omega)\equiv
(\omega^2-m^2)\rho(\omega)$, and verify that when $\omega\to m$,
$F(\omega)$ satisfies
\beq
\frac{ {\nrm d}F}{{\nrm d}\omega}\approx
\frac{(\zeta-3)e^2}{8\pi^2}\,
\frac{2m\,F(\omega)}{\omega^2-m^2}.
\eeq
The solution of this equation gives the behaviour of
$\rho(\omega)$ for $\omega\to m$:
\beq
\rho(\omega)\propto\frac{1}{\omega^2-m^2}\left(
\frac{\omega^2-m^2}{4m^2}\right)^{\frac{(\zeta-3)e^2}{8\pi^2}}.
\eeq
This is the correct behaviour, as obtained by a variety of other
methods (see \cite{LBrown} and Refs. therein).

An essential step in the previous calculation is the cancelation of the
singularity at $s^2=\omega^2$ of the integrand in the r.h.s. of
eq.~(\ref{r1})  by the
phase space factor $(\omega^2-s^2)/\omega^2$ contained in ${\nrm
Im}\,\tilde\Sigma(\omega;s)$, eq.~(\ref{3+1}).
 This phase space factor is very much dependent upon the dimension.
Recall that, in 2+1 dimensions,  it is simply $\sim1/\omega$
(see eq.~(\ref{PhS})), so that in this case the
imaginary part of the one-loop self-energy does not vanish
at threshold; we have indeed
\beq\label{imsig1}
{\nrm Im}\,\tilde\Sigma(\omega;s)\,=\,\pi \alpha\,\frac
{ \omega^2+s^2}{\omega}\,\theta(\omega-s)\,.\eeq
It follows that the integrand in the r.h.s. of eq.~(\ref{r1})
is singular as $s\to \omega$, suggesting that that eq.~(\ref{r1})
is ill defined as written.  However, according to
eq.~(\ref{tilsig}), ${\nrm Re}\,\tilde
\Sigma(\omega;s)$ is also logarithmically divergent when
evaluated for
$s\to \omega$, and it turns out that  this divergence
 precisely cancels that of the r.h.s. To see this,
we write the real part of the one-loop self-energy as
\beq\label{specsig}
{\nrm Re}\, \tilde \Sigma (\omega;m)\,=\,\zeta\,\alpha\,m\,+
\,2\alpha\,\frac{m}{\omega}\int_0^\omega {\nrm d}s\,\frac{\omega^2 +
s^2} {m^2 - s^2}\,.\eeq
The last integral diverges at its upper limit when  $m\to \omega$.
When it is added to the integral in the r.h.s. of eq.~(\ref{r1}),
 the limit $m\to \omega$ becomes well-defined. The equation for
$\rho$ takes then the form
\beq\label{eq}
\rho(\omega)\,\Bigl(\omega^2 - m^2 - \zeta\,\alpha\,\omega \Bigr)
\,=\,2\alpha \int_0^{\omega}{\nrm d}s\,\frac {\omega^2 +s^2}
{{\omega}^2-s^2}\,\Bigl(\rho(\omega) - {\rho(s)}\Bigr)\,.\eeq
This is our main equation. It has a number of properties which
are worth emphasizing.

\noindent i) The integral over $\omega$ in the r.h.s.
 now extend over a limited range of values of $\omega$. This makes it
 well suited to study the behaviour of $\rho(\omega)$ near
threshold.

\noindent ii) It is a linear
integral equation for the spectral density. Being also homogeneous, it
determines
$\rho$ only up to a constant factor, which in principle is fixed by the
inhomogeneous equations (\ref{DS2}) or (\ref{DS3}).
It can be verified in particular that any normalisable solution
of the integral equation (\ref{DS2})
satisfies the sum rule
\beq\label{norm}
\int_0^\infty {\nrm d}\omega\,\rho(\omega)\,=\,1\,.\eeq

\noindent iii) In the limit $\alpha\to 0$, eq.~(\ref{eq})
admits as a solution  the free spectral function,
$\rho(\omega)=2m\delta(\omega^2-m^2)$.

 \noindent iv) The asymptotic form of $\rho(\omega)$ for $\omega \gg m$
has been obtained from perturbation theory
(recall eq.~(\ref{asymp0})):
\beq\label{asymp}
\rho(\omega)\sim -2\,\frac{\alpha}{\omega^2}.
\eeq
This is consistent with
the integral equation (\ref{eq}). Indeed,
by assuming that the solution $\rho(\omega)$ falls off rapidly
enough  in order to be normalisable, one can
 obtain from eq.~(\ref{eq}) the following relation:
\beq
\rho(\omega)\sim -2\frac{\alpha}{\omega^2}
\int_0^\infty{\rm d}s\,\rho(s).
\eeq
When combined with  the sum rule
(\ref{norm}), this equation yields the asymptotic behaviour (\ref{asymp}).
It also shows that $\rho$ must change sign.

\subsection{Solving the integral equation near the mass shell}

{}From the analyticity of the Feynman diagrams (see the discussion
after eq.~(\ref{Sk1})), one expects $\rho(\omega)$ to vanish identically in the
vicinity of $\omega=0$. We  show now that this is indeed the case
for the solution of eqs.~(\ref{DS2}) and (\ref{eq}). Note first
 that  eq.~(\ref{DS2}) is valid for any $k$. Setting
$k=0$, and using $\Sigma(k=0;\omega)=\zeta \alpha \omega$, we obtain
\beq
1\,=\,\int_0^{\infty}{\nrm d}\omega\,\frac{\rho(\omega)}
{\omega^2}\,\Bigl(m^2\,+\,\zeta \alpha \omega \Bigr).\eeq
For this integral to converge,  $\rho(\omega)$ must vanish
sufficiently rapidly  when $\omega \to 0$.
Then, assuming $\rho(\omega)$ to be regular, if it is not
zero it is either increasing or decreasing for $\omega$ small
enough. Assume,
for example, that $\rho$ is increasing, so that it is positive for
small $\omega$. Then, the l.h.s. of the integral equation (\ref{eq}),
 $\approx - m^2 \,\rho(\omega)$,
is negative, while the r.h.s. is positive.
One runs into a similar contradiction if one assumes instead that $\rho$
is decreasing for small $\omega$. The only acceptable possibility is
 that there exists $m^*>0$  such that $\rho(\omega < m^*)=0$.
It may be furthermore
verified on eq.~(\ref{eq}) that $\rho(\omega)$  cannot have
an isolated, $\delta$-type singularity at $\omega=m^*$; that is,
the spectral density is non-vanishing in any upper
vicinity of $m^*$.

Because $\rho(\omega)$ vanishes
when $\omega<m^*$, one can expand the integrand in eq.~(\ref{DS2})
 for small $k$ without generating infrared singularities.
In this way, one obtains sum rules involving higher
 and higher moments of $1/\omega$. For
example, in the Landau gauge $\zeta=0$, we have
\beq\label{SRL1}
1=m^2\int_{m^*}^\infty{\nrm d}\omega\,
\frac{\rho(\omega)}{\omega^2}\,,
\eeq
\beq\label{SRL2}
1=m^2\int_{m^*}^\infty{\nrm d}\omega\,\frac{\rho(\omega)}{\omega^3}
\left(\frac{8}{3}\,\alpha+\frac{m^2}{\omega}\right).
\eeq
Such sum rules suggest that $\rho$ is positive when $\omega\to m^*$.

To study the behaviour of $\rho(\omega)$ for $\omega\simge m^*$,
we divide the  $s$-integration in eq.~(\ref{eq})
in two parts: from 0 to $m^*$, where $\rho(s)=0$, and from
$m^*$ to
$\omega$. After a simple calculation, the integral equation is
rewritten as
\beq\label{eq1}
\rho(\omega)\,\biggl(\omega^2 - m^2 - \zeta\alpha\,\omega
+2\alpha\, m^* -
2\alpha\,\omega\,\ln\frac{\omega+m^*}{\omega-m^*}\biggr)
 \nonumber\\
\qquad\qquad
=\,2\alpha\int_{m^*}^{\omega}{\nrm d}s\,\frac {\omega^2 +s^2}
{{\omega}^2-s^2}\Bigl(\rho(\omega) -
{\rho(s)}\Bigr)\,.\qquad\eeq Assume that
$\rho(\omega)$ is positive near the threshold,
in conformity with the sum rules above.  As $\omega\to m^*$,
the l.h.s. is dominated by the singularity of the
 logarithmic term,
\beq\label{lhs}
{\nrm l.h.s.} \approx -
2\alpha\,m^*\,\rho(\omega)\,\ln\,\frac{2m^*}{\omega-m^*} ,\,\eeq
which is negative  (and  gauge-independent).
The r.h.s. must be negative as well, and this
requires $\rho(\omega)$ to be decreasing. Since $\rho$
is positive and decreasing, it cannot vanish at $m^*$: $\rho(m^*) > 0$.
In fact, $\rho$ is {\it divergent} at  threshold, for,
if  $\rho(m^*)$ were finite, the integral in the r.h.s.
of eq.~(\ref{eq1}) would vanish
as $\omega\to m^*$,  while the l.h.s.,
eq.~(\ref{lhs}), would be divergent.
It follows that, close to the threshold,  the integral in the r.h.s.
of eq.~(\ref{eq1}) is dominated by the singularity of $\rho(s)$
as $s\to m^*$, so that we may approximate
\beq\label{rhs}
{\nrm r.h.s.}\approx -\,2\alpha \,\frac{m^*}{\omega-m^*}\,
\int_{m^*}^{\omega}{\nrm d}s\,\rho(s)\,.\eeq
In writing this equation, we have neglected
 $\rho(\omega)$  (which is finite as long as $\omega > m^*$) next to
$\rho(s)$, and we have made the appropriate
replacements, e.g., $\omega+m^*\to 2
m^*$. From eqs.~(\ref{lhs}) and (\ref{rhs}), we obtain the following
approximate form of the integral
 equation:
\beq\label{limit}
x\,\rho(x)\,\ln \,\frac{1}{x}\,=\,\int_0^x {\nrm d}y\,\rho(y)\,,\eeq
with the notation  $x\equiv (\omega-m^*)/2m^*$ and the
obvious identification $\rho(x)\equiv \rho (\omega=m^* + 2m^* x)$. Note that
$\alpha$ and all the other parameters (namely, $m$ and $\zeta$)
 have dropped from this equation. Its solution is easily found as
\beq\label{rhox}
\rho(x)\,=\,\frac{Z}{2m^*}\,\frac{1}{x\,(\ln x)^2}\, ,\eeq
where $Z$ is a  dimensionless constant.
As  expected, this is divergent as
$x\to 0_+$, but the divergence is integrable, as required for a solution of
the integral equation. In the original variable,
\beq\label{rhop}
\rho(\omega)\,\approx
\,\frac {Z\,\theta (\omega - m^*)}{(\omega-m^*)
\left (\ln\,{{\omega-m^*}\over {2m^*}}\right )^2}\,,
\,\,\,\,\qquad  \omega \to m^*.\eeq
 This is our main result.

Near the mass-shell, the corresponding Minkowski propagator is
($\omega >0$)
\beq\label{Sp}
\tilde S(\omega)\equiv
\int_0^{\infty}{\nrm d}s\,\frac{\rho(s)}
{s^2 - \omega^2 - i\epsilon}\,\approx\,
\frac{Z}{ (\omega^2 - m^{*2})
\ln\,\frac{m^* - \omega - i\epsilon}{2 m^*}}\,.\eeq
This is qualitatively different from the one-loop result
discussed in Sect. 3.1. The inverse propagator
$\tilde S^{-1}(\omega)$ vanishes at
 the mass shell, but its derivative is divergent
there, a situation which is reminiscent of that
 in four dimensions\cite{LBrown}.
Thus, the integral equation (\ref{DS1}) provides, in (2+1)-dimensional SQED,
a mass-shell behaviour for the charged particle
propagator which is analogous to that obtained at the one loop level
in  (3+1)-dimensional SQED.
In particular, the  self-energy corresponding to the propagator
(\ref{Sp}) behaves near the mass-shell as
\beq\label{SEF}
\tilde\Sigma (\omega)\propto
(\omega^2 - m^{*2})
\ln\,\frac{m^* - \omega - i\epsilon}{2 m^*}\,.\eeq
It contains a factor $\omega^2- m^{*2}$
(compare, in this respect, with eqs.~(\ref{imsig})
and (\ref{3+1})) which makes the imaginary part
of (\ref{SEF}) vanish and change sign at the mass-shell.

 According to eq.~(\ref{Scut}),
the long-range behaviour of the screening function is determined
by the spectral density near threshold.
With $\rho(\omega)$ given by eq.~(\ref{rhop}), we get
 $S(x) \sim f(x)\,{\nrm e}^{-m^* x}$, so that $m^*$ plays
the role of the screening mass. Note that if the integral equation
(\ref{limit}) specifies the correct mass-shell behaviour, it leaves
$m^*$ arbitrary.  In principle,  the value of $m^*$ could be obtained by
solving the integral equation (\ref{DS2})
 for $\rho(\omega)$ for {\it all} the values of $\omega$.
But the approximations underlying this equation are
only valid at small momenta.

Now, if we cannot specify  the value of $m^*$ anyfurther, we can
verify that it is independent of the choice
 of the gauge  parameter. To this aim, consider the
gauge-dependent contribution to the scalar self-energy,
as determined by eq.~(\ref{zetasig}).
This can be written in terms of the spectral density as
\beq\label{zeta2}
\delta\Sigma(k)= \, e^2T\, S^{-1}({\nbf k})
\int {\nrm d}\omega\,\rho(\omega)\,
\int \frac{{\nrm d }^3{ q}} {(2\pi)^3}
\,\frac{{\nbf q\cdot(2k+q)}}{(q^2+\lambda^2)^2} \,
\frac{1}{\omega^2+ ({\nbf k+q})^2}\,,\eeq
and  is readily evaluated, with the result
\beq\label{zeta3}
\delta\Sigma(k)= \alpha\, S^{-1}({\nbf k})
\int {\nrm d}\omega\,\frac{\omega\rho(\omega)}
{(\omega+\lambda)^2+ k^2}\,.\eeq
 If we were to set $\lambda =0$, we
would obtain a non-vanishing contribution on the mass-shell:
\beq\label{zeta4}
\delta\Sigma(k)= \alpha\, S^{-1}({\nbf k})
\int {\nrm d}\omega\,\frac{\omega\rho(\omega)}
{\omega^2+ k^2}\,\longrightarrow_{k\to im^*}\,\alpha\, m^*\,.\eeq
When taking the limit $k\to im^*$ in the above equation,
we use the fact that, close to the mass-shell, the integral is
dominated by the singularity of the spectral density at
$\omega=m^*$. In order to get rid of the unwanted
contribution (\ref{zeta4}), which would make the mass-shell
gauge-dependent, we are thus led to keep $\lambda\ne 0$ before
taking the on-shell limit (see the discussion after eq.~(\ref{zetasig})).
 With this procedure, the coefficient
$Z$ in eq.~(\ref{rhop}) is gauge dependent, and
also $\lambda$-dependent, as we explain now.
Consider the equation $S^{-1}=S_L^{-1}+\zeta\delta\Sigma$ (with
$S_L$ the propagator in the Landau gauge) in the vicinity of
the mass-shell, where the approximate form (\ref{Sp})
holds (with $Z_L$ replacing $Z$ in the case of $S_L$).
By using eq.~(\ref{zeta3}) for $\delta\Sigma(k\to im^*)$,
 one derives
\beq\label{Z}
\frac{1}{Z}\,=\,\frac{1}{Z_L}\,+\,\frac{\zeta\alpha}{Z}
\int_{m^*}^\infty {\nrm d}\omega\,\frac{\omega\rho(\omega)}
{(\omega+\lambda)^2 -m^{*2}}\,.\eeq
As $\lambda \to 0$, the integral in the r.h.s. is essentially
the on-shell propagator, so that it is divergent. Thus,
for $\lambda$ small enough, the integral is dominated by
the singularity of $\rho(\omega)$ as $\omega\to m^*$, where
the approximation (\ref{rhop}) can be used. One deduces
that, as $\lambda \to 0$, eq.~(\ref{Z}) takes the form
\beq\label{Z1}
\frac{1}{Z}\,\approx\,\frac{1}{Z_L}\,+\,\frac{\zeta\alpha}{2
\lambda\ln(2m^*/\lambda)}.\eeq
This equation determines the dependence of $Z$ upon $\zeta$
and $\lambda$, in the limit $\lambda \to 0$. We recall here
that $Z_L$ is trivially independent of  $\zeta$
and $\lambda$, since it is determined by eq.~(\ref{DS2})
with $\zeta =0$. According to sum rules like (\ref{SRL1})--(\ref{SRL2}),
we expect $Z_L$ to be positive.

\subsection{The case of a non-vanishing magnetic mass}

For the sake of comparaison with previous computations,
in particular those presented in section 3.1,
it is instructive to consider the  integral equation for the spectral density
 in the presence of an infrared regulator $\lambda \ll m$.  Since
the question of gauge invariance is not an issue here, we use a
single regulator, in contrast to what we did earlier in section 4.1.

With $\lambda \ne 0$, the real and imaginary parts of
 $\tilde\Sigma_\lambda(\omega;s)$ are obtained from
 eq.~(\ref{siglam}):
\beq\label{resig}
{\nrm Re}\,\tilde\Sigma_\lambda(\omega; \omega)
&=& 2\alpha \omega\,\ln\frac{2\omega}{\lambda}
\,-\,\alpha \,\omega\,,\nonumber\\
{\nrm Im}\,\tilde\Sigma_\lambda(\omega;s)&=&\pi \alpha\,\frac
{\omega^2 + s^2}{\omega}\,\theta\Bigl(\omega-(s +\lambda)\Bigr)
-(\zeta-1)\frac{\pi\alpha}{2}\,\frac{s(\omega^2-s^2)}{\omega}\,
\delta\Bigl(\omega-(s +\lambda)\Bigr)\,.\nonumber\\ \eeq
Then, the  homogeneous  integral  equation takes the form
\beq\label{eqlam}
\rho(\omega)\left(\omega^2 - m^2 +\alpha \omega
 -2\alpha \omega \ln\frac{2\omega}{\lambda}
\right)=(\zeta-1)\alpha \omega \rho(\omega-\lambda)-
2\alpha \int_0^{\omega-\lambda}{\nrm d}s\,\frac {\omega^2 +s^2}
{{\omega}^2-s^2}{\rho(s)}\,.\nonumber\\ \eeq
Note the following differences with respect to the case $\lambda =0$:
 {\it i)} ${\nrm Re}\,\tilde\Sigma_\lambda(\omega;\omega)$ is finite;
 {\it ii)}
the $s$-integration in eq.~(\ref{eqlam}) is now restricted
to $s \le \omega -\lambda$, thus avoiding the singularity of the integrand
at $s=\omega$; {\it iii)} the gauge term is proportional to
 $\rho(\omega-\lambda)$ and reflects
 the spurious pole at $k=i(m+\lambda)$ arising in the gauge
piece of the self-energy (\ref{siglam}).

{}From eq.~(\ref{eqlam}), it is easy to establish that the mass-shell
 corresponds to a $\delta$-type singularity, that is, to a simple
pole in the corresponding propagator. To see this,
consider eq.~(\ref{eqlam}) for some $\omega$ satisfying
$m^* < \omega < m^* +\lambda$, where $m^*$ is the mass-shell
position; then, the r.h.s. of the integral equation vanishes
since $\rho(s)=0$  for $s\le \omega-\lambda < m^*$.
The equation becomes, in this momentum range,
\beq\label{eqlam1}
\rho(\omega)\,\left(\omega^2 - m^2 + {\nrm Re}\,
\tilde\Sigma_\lambda(\omega;\omega)\right)
\,=\,0\,,\eeq
and it is solved by $\rho(\omega)\propto \delta(\omega-m^*)$, with $m^*$
determined by
\beq\label{self}
m^{*2}\,=\,m^2\,+\, {\nrm Re}\,\tilde\Sigma_\lambda(m^*;m^*)\,.\eeq
Since ${\nrm Re}\,\tilde\Sigma_\lambda(\omega; \omega)$,
eq.~(\ref{resig}),  is gauge independent, so is $m^*$,
 the solution of the above equation.
Eq.~(\ref{self}) is identical to eq.~(\ref{Rebhan}),
obtained after a partial resummation of
the scalar propagator which  amounts to the replacement of the
leading order mass $m$ by the exact mass $m_D$,  that is,
by using the propagator
\beq\label{SREB} S(k)\,\simeq\,\frac{1}{k^2+m_D^2}\,,\eeq
in the calculation of $\Sigma(k)$. The fact that vertex
corrections do not seem to play any role in the
determination of the mass shell is, strictly speaking, illusory. In fact,
the true propagator in the vicinity of the pole is not (\ref{SREB}),
but rather $z/(k^2+m_D^2)$. The residue $z$ may differ significantly
from  unity; in principle, it could be determined if we were able
to solve the full integral equation (\ref{eqlam}).
However, because of the Ward identity, the residue
enters also the vertex correction, in such a way that it cancels
against that of the propagator when the self-energy is
computed at the mass-shell. It is therefore not needed to determine
$m^*$, but it enters as a preexponential factor in the asymptotic
form of the screening function. We see now that the
 preexponential factor in eq.~(\ref{res}) is not consistently determined.

\setcounter{equation}{0}
\section{Conclusions}

We have shown that the corrections to the Debye mass in high temperature
non abelian gauge
theories can be analyzed as corrections to the mass-shell of
 particles coupled to massless magnetic modes in a
 2+1 dimensional effective theory. If
one attempts to calculate perturbatively
the Debye mass  within this effective theory,
 one encounters power-like infrared divergences
which signal a breakdown of the perturbative expansion.
The possible existence of  a
magnetic mass in QCD does not
remove the essentially non perturbative character
of the corrections.

We have shown that similar mass-shell singularities occur in the
evaluation of  the scalar propagator in scalar
electrodynamics. For this case, we have presented a non
perturbative approach which allows for a complete description of the
mass-shell behaviour. The mass-shell singularity,
 which is a simple pole in leading order, turns
into a branch point as a result of the coupling of the scalar particles to
an arbitrary number of soft photons. The propagator near the branch point
exibits a behaviour which is reminiscent of that of the one-loop propagator
in 3+1 dimensional electrodynamics.
The location of the branch point, which plays
the role of the screening mass in the thermal problem, is shown to be gauge
independent. However, its precise value is left undetermined by our present
approach.

The calculation that we have performed for SQED
suggests that a similar solution
may exist for QCD as well. If a magnetic mass exists,
the mass-shell remains a pole whose exact location
is determined by the
integral equation that we have derived, assuming that
this equation applies also to QCD; the
correction to the Debye mass thus obtained is then
 identical to that calculated by Rebhan\cite{Rebhan94}.
 Finally, we believe that the present analysis should also shed
light on another longstanding problem in finite temperature
field theory, that
is, the infrared singularity of the damping rates for thermal particles
(see, e.g., \cite{Rebhan95} and references therein).

 \vspace{1cm}
\noindent
{\bf Acknowledgements}\\
\noindent We thank A. Rebhan for his reading of the manuscript
and for useful comments.

\setcounter{equation}{0}
\vspace*{2cm}
\renewcommand{\theequation}{A.\arabic{equation}}
\appendix{\noindent {\large{\bf Appendix}}}

We verify here that, close to  the mass-shell $k^2 \to - m^2$,
 the two-loop rainbow diagram, Fig. 4.a,
 becomes as important as the one-loop graph of Fig. 2 if
the infrared cut-off $\lambda$ is of the order $g^2T$ or less.

A straightforward application of the Feynman rules for the action
(\ref{Sred}) gives (with $({\nrm d} q) \equiv
 {{\nrm d}^3 q / (2\pi)^3}  $)
\beq\label{Sig2}
\Sigma^{(2)}(k)= - (e^2 T)^2\,
 \int \frac{({\nrm d} q)}{{\nbf q}^2+\lambda^2}\,
\int\frac {({\nrm d} p)}{{\nbf p}^2+\lambda^2}
\,\frac{({\nbf q + 2k})^2}{\left [({\nbf q+k})^2+m^2\right ]^2}
\,\frac{\left({\nbf p+ 2 k +2q}\right)^2}{({\nbf p+ q+k})^2+m^2}\,,
\eeq
where we allow for a photon mass
 $\lambda$ as a convenient IR regularization,
and we use the Feynman gauge $\zeta = 1$ for simplicity. We are interested
here only in the dominant  IR singularity of $\Sigma^{(2)}(k)$ as $k \to \pm
 i m$ and $\lambda \to 0$.
After working out the scalar products in the numerator, we isolate the terms
with the largest number of factors in the denominator. These terms are the
most singular in the double limit
 $k^2 \to - m^2$ and  $\lambda \to 0$. (If either
$k^2 \ne - m^2$ or  $\lambda \ne 0$,
$\Sigma^{(2)}$ is finite.) We denote the corresponding contribution to
 (\ref{Sig2}) by $\Sigma^{(2)}_{IR}(k)$. We have
\beq\label{Sig3}
\Sigma^{(2)}_{IR}(k)= - 16 m^4\,(e^2 T)^2\,
 \int \frac{{\nrm d}_3 q}{{\nbf q}^2+\lambda^2}\,
\int\frac {{\nrm d}_3 p}{{\nbf p}^2+\lambda^2}
\,\frac{1}{\left [({\nbf q+k})^2+m^2\right ]^2}
\, \frac{1}{({\nbf p+ q+k})^2+m^2}\,.\eeq
The terms which have been  neglected in going  from eq.~(\ref{Sig2})
to eq.~(\ref{Sig3}) are, at most, logarithmically divergent in
the double limit mentioned before, and do not matter for the power
counting developed in Sect. 3. As we shall see, the
integral (\ref{Sig3}) is linearly (or, more accurately, linearly
$\times$ logarithmically) divergent in the same limit.

To perform the momentum integrals in eq.~(\ref{Sig3}), we use
the coordinate-space representation of the Coulomb propagators, e.g.
\beq\label{coul}
\frac{1}{k^2+m^2} = \int {\nrm d}^3x\,{\nrm e}^{i{\vec k\cdot \vec
x}}\,\,\frac {{\nrm e}^{-mx}}
{4\pi x},\eeq
(with $x\equiv |{\vec x}|$) and  obtain
\beq\label{Sig4}
\Sigma^{(2)}_{IR}(k)= - \frac{2\alpha^2 m^3}{(2\pi)^2}
 \int {{\nrm d}^3 x}  \int {{\nrm d}^3 y}\,\,
{\nrm e}^{i{\vec k\cdot \vec x}}\,{\nrm e}^{-m|{\vec x - \vec y}|}\,\,
\frac{{\nrm e}^{-\lambda x}}{x}\,
\frac{{\nrm e}^{-(m+\lambda)y}}{y^2}\equiv -2\alpha^2 J(k;m,\lambda),\eeq
where  $\alpha = e^2T/4\pi$.  Once the
angular integrations are done, $J$ is given by
\beq\label{J}
J(k;m,\lambda)=  \frac{2m}{k}
 \int_0^\infty \frac{{\nrm d}x}{x}  \int_0^\infty \frac{{\nrm d}y}{y}
\,{\sin{kx}}\,{{\nrm e}^{-\lambda x}}\,
{\nrm e}^{-(m+\lambda)y}\,F(x,y)\,,\eeq
where
\beq\label{Fxy}
F(x,y)\equiv \Bigl(1+m|x-y|\Bigr)\,{\nrm e}^{-m|x-y|}\,-\,
 \Bigl(1+m(x+y)\Bigr)\,{\nrm e}^{-m(x+y)}\,=\,F_1\,+\,F_2\,,\eeq
with
\beq\label{F12}
F_1(x,y)\equiv m\left\{|x-y|\,{\nrm e}^{-m|x-y|}\,-\,
(x+y)\,{\nrm e}^{-m(x+y)}\right \}\,\\ \nonumber
F_2(x,y)\equiv\,{\nrm e}^{-m|x-y|}\,-\,{\nrm e}^{-m(x+y)}\,.\eeq
The IR divergences occuring in the original momentum integrals
 as $k^2 \to -m^2$ and $\lambda \to 0$ appear now as UV divergences
 of the integrals over $x$ and $y$. The most singular terms
are generated by $F_1(x,y)$. The contribution of $F_2(x,y)$
is, at most, logarithmically divergent. Let $J_1(k;m,\lambda)$ denote the
 contribution of $F_1$ to the integral (\ref{J}). For $\lambda =0$, but
arbitrary $k$ we have a simple expression:
\beq\label{J1}
J_1(k;m,\lambda =0)=  \frac{2m^2}{k^2+m^2}\,\left( \ln \frac{4m^2}
{k^2+m^2}\,-\,1\right)\,.\eeq
As $k^2\to -m^2$, this has a linear $\times$ logarithmic singularity,
as announced. For $\lambda\ne 0$, the expression of $J_1$ is
more complicated.
We give here only the explicit form of the
{\it most singular} term, valid for arbitrary $k$ and $\lambda$:
 \beq\label{JIR}
J_{IR}(k;m,\lambda)=  \frac{m}{ik}\,\left\{\frac{m}{m+\lambda - ik}
 \ln \frac{2m+ \lambda}{m+2\lambda -ik}\,-\,
\frac{m}{m+\lambda +ik}
 \ln \frac{2m+ \lambda}{m+2\lambda +ik}\right\}\,.\eeq
 After inserting this in eq.~(\ref{Sig4}), we
obtain the
dominant (i.e. the most singular) mass-shell behaviour of the  two-loop
diagram
(\ref{Sig2}) as
\beq\label{SIR}
\Sigma^{(2)}_{IR}(k^2 \to -m^2)
 \approx  - {2\alpha^2}\, \frac{m}{\lambda}\,\ln\,\frac{m}{\lambda}\,
\sim  \,\left(\frac{\alpha}{\lambda}\right)\,\Sigma^{(1)}(k^2\to
-m^2)\,,\eeq
(recall eq.~(\ref{siglam})). Thus, for $\lambda$
of the order $e^2T\sim \alpha$ or smaller, and  in the vicinity
of the mass-shell, the two-loop contribution  $\Sigma^{(2)}_{IR}$ is
of the same order in $\alpha$ as the one-loop self-energy.
 A similar conclusion is reached in section 3.3
using power counting.

  \newpage

\begin{center}
{\Large\bf Figure captions}
\end{center}
\vspace*{1cm}
\noindent Figure 1. Analytical structure of the screening function
 $S(k)$ in the complex $k$-plane, showing the branch point at
$k=im^*$.  The contour is that used for the evaluation
of the integral in eq.~(\ref{S1}).

\vspace*{1cm}
\noindent Figure 2. The one-loop  contribution to the
 self energy in the  three-dimensional effective theory.
Full line: scalar (electrostatic) field. Wavy line: transverse gluon.

\vspace*{1cm}
\noindent Figure 3. Approximate representation
of the trajectories followed in the complex
$k$-plane by the poles of the one-loop scalar propagator.
The arrows indicate the flows of the poles as the coupling increases.
With the present choice of gauge ($\zeta=2$), the poles become
real when $\alpha/m \simge 1.45$.

\vspace*{1cm}
\noindent Figure 4. Two-loop contributions to the scalar self-energy
 which contain linear mass-shell divergences. Diagram (a) and (b)
occur both in QCD and in SQED, while diagram (c)  exists only in QCD.

\vspace*{1cm}
\noindent Figure 5. An exemple of a multi-loop contribution to
the self-energy $\Sigma(k)$, exhibiting power-like mass-shell divergences.

\vspace*{1cm}
\noindent Figure 6.  The skeleton diagrams for the self-energy
 of the scalar propagator  in SQED. The dominant infrared
singularities are contained in diagram 6.a.

\vspace*{1cm}
\noindent Figure 7. Diagrammatic representation of the Dyson-Schwinger
equation of SQED in the quenched approximation. Both the internal
scalar propagator (heavy line) and the vertex (heavy blob)
are exact quantities, while the photon propagator (wavy line)
is the bare propagator.

\end{document}